\documentclass[a4paper,11pt]{article}
\pdfoutput=1\usepackage{jcappub}
\usepackage{mathtools}
\usepackage{dsfont}
\graphicspath{{plots/}}

\newcommand{\gs}{g_\star}
\newcommand{\gss}{g_{\star s}}
\newcommand{\Trh}{T_\text{rh}}
\newcommand{\Tbbn}{T_\text{BBN}}
\newcommand{\Tfo}{T_\text{fo}}
\newcommand{\arh}{a_\text{rh}}
\newcommand{\afo}{a_\text{fo}}
\newcommand{\xfo}{x_\text{fo}}
\newcommand{\xrh}{x_\text{rh}}
\newcommand{\sv}{\langle\sigma v \rangle}
\newcommand{\svk}{\langle\sigma_{r \to 2} v^{r-1} \rangle}
\newcommand{\rR}{\rho_R}
\newcommand{\rp}{\rho_\phi}
\title{Unitarity Bound on Dark Matter in Low-temperature Reheating Scenarios}
\author[a]{Nicolás Bernal,}
\author[b]{Partha Konar,}
\author[b]{Sudipta Show}

\affiliation[a]{New York University Abu Dhabi\\
PO Box 129188, Saadiyat Island, Abu Dhabi, United Arab Emirates}
\affiliation[b]{Physical Research Laboratory \\ Ahmedabad - 380009, Gujarat, India}

\emailAdd{nicolas.bernal@nyu.edu}
\emailAdd{konar@prl.res.in}
\emailAdd{sudipta@prl.res.in}

\abstract{Model-independent theoretical upper bound on the thermal dark matter (DM) mass can be derived from the maximum inelastic DM cross-section featuring the whole observed DM abundance. We deploy partial-wave unitarity of the scattering matrix to derive the maximal thermally-averaged cross section for general number-changing processes $r\to 2$ (with $r\ge 2$), which may involve standard model particles or occur solely within the dark sector. The usual upper limit on the DM mass for $s$-wave annihilation is around 130~TeV (1~GeV) for $r=2$ (3), only applies in the case of a freeze-out occurring in the standard cosmological scenario. We consider the effects of two nonstandard cosmological evolutions, characterized by low-temperature reheating: $i)$ a kination-like scenario and $ii)$ an early matter-dominated scenario.
In the first case, early freeze-out strengthens the unitarity bound to a few TeVs for WIMPs; while in the second case, the WIMP DM can be as heavy as $\sim 10^{10}$~GeV due to a large entropy dilution.}

\begin{document} 
\begin{flushright}
\end{flushright}
\maketitle

%%%%%%%%%%%%%%%%%%%%%%%%%%%%%%%%%%%%%%%%%%%%%%%%%%%%%%%%%
\section{Introduction}
%%%%%%%%%%%%%%%%%%%%%%%%%%%%%%%%%%%%%%%%%%%%%%%%%%%%%%%%%
The omnipresence of nonbaryonic dark matter (DM) has been confirmed from celestial observations at different scales, from individual galaxies to clusters of galaxies, and even at the cosmological scale~\cite{Bertone:2016nfn}. Several properties of DM also emerge from such observations, albeit only by exploiting its gravitational interaction.
However, observational evidence has not yet answered the basic questions of the nature, properties, and other interactions of DM. Also, in the absence of any hints of complexity and interactions with and within the dark sector, the general consensus deems DM as particles of a fundamental nature beyond our known form of matter profoundly established in the standard model (SM) of particle physics.
Such a DM is required to be electrically neutral, stable at the cosmological time scale, and non-relativistic at the time of matter-radiation equality to permit structure formation. Furthermore, the observation of the cosmic microwave background (CMB) established that DM holds a 27\% share of the total energy budget of the present universe, corresponding to a relic density $\Omega h^2 \simeq 0.12$~\cite{Planck:2018vyg, Drees:2018hzm}.  

With a decade-old theoretical development in modeling different particle DM candidates, the extensive possibility of a wide mass range of viable DM candidates has been opened~\cite{Rubakov:2017xzr}. This, in turn, pushed the community to develop ingenious approaches for the next generation of DM experiments, covering ultralight to very heavy ones. Among this wide range of DM mass, the model-independent lower limit of around $10^{-22}$~eV comes from the de Broglie wavelength of bosonic DM that can be confined within a dwarf galaxy~\cite{Hu:2000ke, Hui:2016ltb, Nori:2018pka}. This paradigm of ultralight fuzzy DM candidates has received significant attention lately due to its ability to alleviate tensions between anomalies ranging from theoretical modeling and simulation structure formation and vast astrophysical data. 
Further refinement of the DM mass bounds is influenced by the specific properties and nature of DM. Notably, the lower limit for the DM mass is more constrained in the case of fermionic DM, because of the Pauli exclusion principle. Furthermore, observations of the Lyman-$\alpha$ forest play a crucial role in establishing constraints on DM. These observations set a lower limit of around 5.3~keV~\cite{Irsic:2017ixq} for warm DM by constraining its free streaming length.
In contrast, a model-independent upper limit of the DM mass can be derived in the range of $10^3$ solar mass from the stability of stellar clusters in galaxies~\cite{Moore:1993sv, Carr:1997cn}. Once again, observations of the Lyman-$\alpha$ forest play a role in further constraints from Poisson noise~\cite{Afshordi:2003zb}.

The consideration of a specific DM production paradigm in the early stage of the universe may further constrain the mass range for a viable DM candidate. For instance, the number-changing pair annihilation of DM to SM particles determines its present mass density, where it maintains the chemical and kinetic equilibrium with the thermal soup in the early universe. Interestingly, the requirement of the unitarity of the $S$-matrix sets a model-independent upper bound on the DM mass for this scenario~\cite{Griest:1989wd, Hui:2001wy, Bhatia:2020itt}. The implication of unitarity offers the maximum inelastic cross section, which fixes the minimum number density of the frozen-out DM. Using this number density, one can establish the maximum allowed DM mass by fulfilling its observed relic density. In the theories of DM with long-range forces, DM bound states can form which lessen the unitarity bound than the picture without bound states, since the formation of bound states reduces the effective inelastic annihilation rate~\cite{vonHarling:2014kha, Baldes:2017gzw, Smirnov:2019ngs}. In addition, the dark sector with particle-antiparticle asymmetry enforces a nonzero equilibrium chemical potential for DM, which further constrains the unitarity limits by demanding an increased effective DM number density at the time of freeze-out~\cite{vonHarling:2014kha, Ghosh:2020lma}. Furthermore, different indirect searches for DM may put a lower limit on the DM mass for some specified scenarios. A strong model-independent lower bound for thermal DM that is annihilating to visible states through an $s$-wave process is about 20~GeV~\cite{Leane:2018kjk}. In addition, a more restrictive lower limit has recently been found. It has been shown that the lower bound is 200~GeV, considering H.E.S.S. and other updated observational data~\cite{Dutta:2022wdi}.

In particular, all the DM scenarios mentioned so far pay attention to the $2 \to 2$ number-changing process where a DM pair annihilates into a pair of SM particles, that is, the Weakly Interacting Massive Particle (WIMP) paradigm~\cite{Lee:1977ua, Arcadi:2017kky}.\footnote{Alternatively, one can also have in the final state a DM and a SM particle (semi-annihilations)~\cite{Hambye:2008bq, Hambye:2009fg, DEramo:2010keq, Belanger:2012zr, Belanger:2014bga}, or in the initial state a DM and another particle of the dark sector (coannihilations)~\cite{Griest:1990kh}.}
Moreover, it is not necessary that the number-changing processes involve SM particles, so they may also occur within the dark sector. The minimalist realization of this scenario is the $3 \to 2$ process, where this kind of number-changing reaction involves a single DM species. In general, such processes arise in DM theories with new sizable self-interactions, and in several contexts as self-interacting DM~\cite{Carlson:1992fn, Pappadopulo:2016pkp, Farina:2016llk}, the Strongly Interacting Massive Particle (SIMP) paradigm~\cite{Hochberg:2014dra, Choi:2015bya, Bernal:2015bla, Bernal:2015lbl, Ko:2014nha, Choi:2017mkk, Chu:2017msm, Bernal:2015ova, Yamanaka:2014pva, Hochberg:2014kqa, Lee:2015gsa, Hansen:2015yaa, Bernal:2015xba, Heikinheimo:2016yds, Bernal:2017mqb, Heikinheimo:2017ofk, Bernal:2018ins, Bernal:2018hjm}, or even the ELastically DEcoupling Relic (ELDER) scenario~\cite{Kuflik:2015isi, Kuflik:2017iqs}.
 
Now, the question arises: What would be the consequences of the unitarity of the $S$-matrix for a general DM number-changing process of the kind $r\to n$ with $r> n \ge 2$? Here, we may focus on a subset of the reaction with the type $r \to 2$ instead of the generalized $r\to n$ interaction to avoid the complexity of handling the partial-wave decomposition for the $r$-body initial state. However, the partial-wave analysis for a two-body initial state is simple. The thermally averaged cross section $\langle\sigma_{r \to 2} v^{r-1}\rangle$ can be easily obtained from $\langle\sigma_{2 \to r} v\rangle$ utilizing the fact that both cross sections are related in thermal equilibrium. The implication of unitarity helps to calculate the maximum inelastic cross section for the $2\to r$ process once the total cross section, calculated using the optical theorem and the elastic scattering cross section for the $2 \to 2$ process are known.

It is essential to mention that the early history of the universe plays a crucial role in the genesis of DM, since the decoupling of thermal DM occurred at that time. Generally, the studies of DM consider the standard cosmological picture in which the radiation energy density is assumed to dominate the energy budget before the Big Bang nucleosynthesis (BBN). However, there is no direct evidence for the energy content at very high temperatures. Therefore, it is vital to look at the effects of modified cosmology on the production of DM. In recent times, the evolution of the DM in the period of non-standard expansion usually triggered by the decay of a long-lived massive particle~\cite{Giudice:2000ex, Fornengo:2002db, Pallis:2004yy, Gelmini:2006pw, Drees:2006vh, Yaguna:2011ei, Roszkowski:2014lga, Harigaya:2014waa, Drees:2017iod, Bernal:2018ins, Bernal:2018kcw, Cosme:2020mck, Ghosh:2021wrk, Arias:2021rer, Bernal:2022wck, Bhattiprolu:2022sdd, Haque:2023yra, Ghosh:2023tyz, Silva-Malpartida:2023yks, Arias:2023wyg, Das:2023owa} or by Hawking evaporation of primordial black holes~\cite{Green:1999yh, Khlopov:2004tn, Dai:2009hx, Fujita:2014hha, Allahverdi:2017sks, Lennon:2017tqq, Morrison:2018xla, Hooper:2019gtx, Chaudhuri:2020wjo, Masina:2020xhk, Baldes:2020nuv, Gondolo:2020uqv, Bernal:2020kse, Bernal:2020ili, Bernal:2020bjf, Bernal:2021akf, Cheek:2021odj, Cheek:2021cfe, Bernal:2021yyb, Bernal:2021bbv, Bernal:2022oha, Cheek:2022dbx, Mazde:2022sdx, Cheek:2022mmy} is receiving increasing attention.\footnote{For studies on baryogenesis with a low reheating temperature or during an early matter-dominated phase, see refs.~\cite{Davidson:2000dw, Giudice:2000ex, Allahverdi:2010im, Beniwal:2017eik, Allahverdi:2017edd, Konar:2020vuu} and~\cite{Bernal:2017zvx, Chen:2019etb, Bernal:2022pue, Chakraborty:2022gob}, respectively. Furthermore, the production of primordial gravitational waves in scenarios with an early matter era has recently received particular attention~\cite{Assadullahi:2009nf, Durrer:2011bi, Alabidi:2013lya, DEramo:2019tit, Bernal:2019lpc, Figueroa:2019paj, Bernal:2020ywq}.}
All such studies point towards the fact that nonstandard cosmology alters the value of the thermally averaged cross section needed to satisfy the observed relic of DM. Such a modification in the thermally averaged cross section may also change the unitarity mass bound of the DM. Recently, the impact of early matter domination on unitarity limits has been studied~\cite{Bhatia:2020itt}.

The present work explores how such bound modifies in the context of two different non-standard cosmological pictures. One is the kination-like scenario in which a species, $\phi$, dominates the energy density in the pre-BBN era. Here, the energy density of $\phi$ maintains the following redshift behavior, $\rho_\phi\propto a^{-(4+n)}$ with $n>0$ where $a$ is the scale factor~\cite{DEramo:2017gpl}. Another is late-time reheating, where delayed reheating occurs after inflation due to the suppressed interaction of the inflaton field~\cite{Giudice:2000ex}. In general, it can be realized as a scenario of fast expansion with entropy injection. We demonstrate that the kination-like scenario restricts the unitarity limits compared to the standard case. Since one needs a larger cross section than a standard picture due to the early freeze-out of DM. However, the picture is interestingly opposite in the case of late-time reheating. Although the DM decouples earlier because of the fast expansion, the unitarity bounds are relaxed here because of the entropy injection after the decoupling of DM from the thermal soup.

This article is decorated as follows. In Sections~\ref{sec2} and~\ref{sec3}, we present the detailed derivation of the maximum thermally-averaged cross section allowed by the unitarity of the $S$-matrix. We discuss two different non-standard cosmological pictures: kination-like and late-time reheating in Section~\ref{sec4}. Section~\ref{sec5} shows the analytical expressions for freeze-out and cross sections for the radiation-dominated universe and the mentioned modified cosmologies, and we also demonstrate our results. Finally, we summarize our findings in Section~\ref{sec6}.

%%%%%%%%%%%%%%%%%%%%%%%%%%%%%%%%%%%%%%%%%%%%%%%%%%%%%%%%%
\section{\boldmath $S$-matrix: Unitarity and its Consequences} \label{sec2}
%%%%%%%%%%%%%%%%%%%%%%%%%%%%%%%%%%%%%%%%%%%%%%%%%%%%%%%%%
$S$-matrix theory gives the information about the probability amplitude of various outcomes of a scattering process~\cite{Weinberg:1995mt, Schwartz:2014sze}. Interestingly, the unitarity of the $S$-matrix dictates the conservation of the probability of interaction. Although the unitarity of the $S$-matrix appears to be apparently simple, it has some remarkable implications. We would recollect some consequence of it in the case of DM annihilation, since our primary interest is to put a bound for DM scattering. Here, we consider the multiparticle momentum eigenstate normalization convention~\cite{Hui:2001wy} and closely follow the derivation in Ref.~\cite{Bhatia:2020itt}. One of the outstanding outcomes of the $S$-matrix is the optical theorem, which relates the imaginary part of the scattering amplitude ($\mathcal{A}_{\beta\beta}$) with the total cross section ($\sigma_\text{total}$) and can be expressed in the center-of-mass frame as
\begin{equation} \label{opt_theorem}
	\text{Im}~\mathcal{A}_{\beta\beta} = 2\, |\vec{k}|\, E_\text{cm}\, \sigma_\text{total}\,,
\end{equation}
with $\beta$ describes the two-particle initial state. Here, $|\vec{k}|$ and $E_\text{cm}$ represent the magnitude of the three momenta of each particle in the initial state and the total energy of the initial state in the CM frame, respectively. The state $\alpha$ is delineated by the three momenta, the $z$ component of the spin (or helicity) of the individual particles along with all other internal quantum numbers, and $\sigma_\text{tot} \equiv \sum_\gamma \sigma_{\beta \to \gamma}$ where $\gamma$ corresponds to the possible final states. It is important to mention that our focus will be on the interaction of nonidentical scalar particles for simplicity, and generalization to other spin can be done following Refs.~\cite{Hui:2001wy, Weinberg:1995mt, Schwartz:2014sze}. Here, we would work on a general basis state instead of the basis of momentum eigenstates, which can be done as a result of rotational invariance.

One needs to know the amplitude first to calculate the cross section, and to do so, let us focus on the operator of the $S$-matrix ($\mathcal{S})$. It is obvious that the initial and final states are identical in the absence of any interaction. Therefore, the operator $\mathcal{S}$ is simply the identity $\mathds{1}$, as nothing has happened. If an interaction occurs, one needs to subtract the identity from $\mathcal{S}$, where all the information about the interaction is encoded in the nontrivial part $\mathcal{S}-\mathds{1}$. Now, we can express the operator of the $S$-matrix using the definition of amplitude as
\begin{equation} \label{S_matrix_22}
    \mathcal{S}_{\gamma\beta} = \delta(\gamma - \beta) + {(2\pi)}^4\, \delta^{(4)}(k_\gamma - k_\beta)\, (i\, \mathcal{A}_{\gamma\beta})\,,
\end{equation}
where $k_\beta$ and $k_\gamma$ represent total four momenta.
Furthermore, the $S$-matrix can be expressed as a function of the quantum number of the total angular momentum $l$, the $z$ component of the total angular momentum quantum number $\mathtt{m}$ and the total energy $E_\text{tot}$. Using the relation of Eq.~\eqref{S_matrix_22}, the general expression of the matrix element for 2-to-2 scatterings $|\vec{k}_1, \vec{k}_2, \mathtt{q} \rangle \to |\vec{k}_1', \vec{k}_2', \mathtt{q}'\rangle$ can be written as\footnote{Here we have used the fact that that $\mathcal{S}$ commutes with the generators of rotations and so the matrix element is independent of $\mathtt{m}$.}
\begin{equation} \label{Mat_ele_22}
    \mathcal{A}_{\mathtt{q}' \mathtt{q}} = -i\, \frac{16\pi^2\, E_{\text{tot}}}{\sqrt{|\vec{k}_1| |\vec{k}_1'|}} \sum_{l,\mathtt{m}} Y_{l \mathtt{m}}(\hat{k}_1')\, Y_{l \mathtt{m}}(\hat{k}_1)^* \left(\mathcal{S}_{\mathtt{q}' \mathtt{q}}(l,E_{\text{tot}}) - \delta_{\mathtt{q}' \mathtt{q}}\right),
\end{equation}
where $\mathtt{q}$ and $\mathtt{q}'$ represent the channel indices for the two particle species. Here, $Y_{l \mathtt{m}}(\hat{k}_1')$ and $Y_{l \mathtt{m}}(\hat{k}_1)$ correspond to the spherical harmonics in the direction of the momentum unit vector $\hat{k}_1'$ and $\hat{k}_1$. Then, the total cross section, the sum of the elastic and inelastic cross sections, in a given channel can be obtained with the help of the optical theorem in Eq.~\eqref{opt_theorem} and the expression of the matrix element in Eq.~\eqref{Mat_ele_22}
\begin{equation} \label{tot_non_iden}
	\sigma_\text{total} = \sum_l \frac{\pi}{|\vec{k}_1|^2} (2 l + 1)\, 2\, \text{Re}\left(1 - \mathcal{S}_{\mathtt{q} \mathtt{q}}(l,E_{\text{tot}})\right).
\end{equation}
Then, we obtain the elastic scattering cross section $\sigma_\text{el}$ in the CM frame from the channel $\mathtt{q}$ after performing the integration over $d\Omega(\hat{k}_1')$ while choosing $\hat{k}_1$ along the $z$ direction
\begin{equation} \label{elas_non_iden}
	\sigma_\text{el} = \sum_l\frac{\pi}{|\vec{k}_1|^2}\, (2l+1)\, |(\mathcal{S}_{\mathtt{q} \mathtt{q}}(l,E_{\text{tot}})-1)|^2.
\end{equation}
Now, one can easily obtain the inelastic cross section by subtracting the elastic in Eq.~\eqref{elas_non_iden} from the total cross sections in Eq.~\eqref{tot_non_iden}
\begin{equation} \label{inelas_non_iden}
	\sigma_\text{inel} = \sum_l \frac{\pi}{|\vec{k}_1|^2} (2 l + 1) (1 - |\mathcal{S}_{\mathtt{q} \mathtt{q}}(l,E_{\text{tot}})|^2)\,.
\end{equation}
Finally, we obtain the upper limits on the total inelastic cross section taking into account that $|\mathcal{S}_{\mathtt{q} \mathtt{q}}(l, E_{\text{tot}})|^2\ge 0$
\begin{equation} \label{inelas_non_iden_max}
	\sigma_\text{inel} \le \sum_l \frac{\pi}{|\vec{k}_1|^2} (2 l + 1)\,.
\end{equation}
The above expression can also be expressed in terms of velocity using the fact that, in the nonrelativistic limit, the momentum can be expressed as $|\vec{k}_1| \simeq m \, v/2$, where $v$ refers to the relative velocity of the annihilating particles. Finally, we note that for a collision among identical particles, an additional multiplicative factor of $2$ is required in the cross section to avoid double counting~\cite{Weinberg:1995mt}.

%%%%%%%%%%%%%%%%%%%%%%%%%%%%%%%%%%%%%%%%%%%%%%%%%%%%%%%%%
\section{Dark Matter Annihilation and Unitarity Bound} \label{sec3}
%%%%%%%%%%%%%%%%%%%%%%%%%%%%%%%%%%%%%%%%%%%%%%%%%%%%%%%%%
We have obtained the upper bound on the inelastic cross-section, and now we are interested in calculating the unitarity bound on the inelastic reaction rate with the help of the Boltzmann equation for DM. The Boltzmann equation for the number-changing DM annihilation of the $i^\text{th}$ particle in the $r \to 2$ process, where $r\ge 2$, i.e., ($1,\, 2, \cdots,\, i,\, \cdots,\, r \to a,\, b$) in the isotropic and homogeneous expanding early universe, can be manifested as
\begin{equation} \label{BEQ1}
    \frac{dn_i}{dt} + 3 H\, n_i = -\sum_\text{Channels} \Delta n_i\, \big[n_1\, n_2 \cdots n_r\, \langle\sigma_{r\to 2} v^{r - 1}\rangle - n_a\, n_b\, \langle\sigma_{2\to r} v\rangle\big],
\end{equation}
with $\Delta n_i$ being the net change in the $i^\text{th}$ particle number in the reaction, and where the thermally-averaged cross section is given by
\begin{equation} \label{svrt2}
    \svk \equiv \frac{\int d^3 k_1 \cdots d^3 k_r\, f_1^\text{eq} \cdots f_r^\text{eq}\, \sigma_{r\to 2}\, v^{r-1}}{\int d^3k_1 \cdots d^3k_r\, f_1^\text{eq} \cdots f_r^\text{eq}}\,,
\end{equation}
where $\sigma_{r\to2}$ represents the cross section, summed over final spins and averaged over initial spins, for the $r\to2$ process, $v$ is the relative velocity between each particle pair, and $f_i^\text{eq}$ refers to the equilibrium distribution function of the $i^{\text{th}}$ particle species.
More specifically, $n_1^\text{eq}\, n_2^\text{eq} \cdots n_r^\text{eq}\, \svk$ is the equilibrium rate density for the annihilation.
Similarly, one can perform the thermal average to obtain $\langle\sigma_{2\to r}v\rangle$. Using the relation of detailed balance, $n_1^\text{eq}\, n_2^\text{eq} \cdots n_r^\text{eq}\, \svk = n_a^\text{eq}\, n_b^\text{eq}\, \langle\sigma_{2\to r} v\rangle$ (true for any individual process in equilibrium), Eq.~\eqref{BEQ1} boils down to the following simplified form 
\begin{equation} \label{BEQ2}
    \frac{dn_i}{dt} + 3 H\, n_i = -\sum_\text{Channels} \Delta n_i\, n_a^\text{eq}\, n_b^\text{eq}\, \langle\sigma_{2\to r} v\rangle \left[\frac{n_1\, n_2 \cdots n_r}{n_1^\text{eq}\, n_2^\text{eq} \cdots n_r^\text{eq}} - \frac{n_a\, n_b}{n_a^\text{eq}\, n_b^\text{eq}}\right].
\end{equation}
Using the maximum value of the inelastic cross section from Eq.~\eqref{inelas_non_iden_max}, one can calculate the maximal value $\langle\sigma_{2\to r} v\rangle_\text{max}$ for the thermal average integral in Eq.~\eqref{svrt2} for the $2\to r$ process, considering equal masses for all $r+2$ particles participating in the interaction, as\footnote{Although the expression of Eq.~\eqref{MaxCross2tr} is derived considering equal mass for all particles involved in the interaction; it provides the maximum thermally-averaged cross section for WIMPs~\cite{Lee:1977ua, Arcadi:2017kky, Konar:2020wvl} in the case $r=2$ where the mass of the final-state particles can be different from the same of the initial-state particles.}~\cite{Bhatia:2020itt}
\begin{equation} \label{MaxCross2tr}
    \langle\sigma_{2\to r} v\rangle_\text{max} = \sum_l (2 l + 1)\, \frac{4 \sqrt{\pi}}{m^2}\, \sqrt{x}\, e^{-(r - 2) x},
\end{equation}
where $x \equiv m/T$ and $T$ is temperature of the SM bath. It is evident that $\svk_\text{max}$ contains an exponential suppression factor, $e^{-(r-2)x}$, which reveals the expense of phase space to produce each additional particle for $r\ge 3$. Now, we can obtain the maximum thermally averaged rate for the $r \to 2$ process using Eq.~\eqref{MaxCross2tr} and the detailed balance equation, as~\cite{Bhatia:2020itt}
\begin{equation} \label{MaxCrossrt2}
    \svk_\text{max} = \sum_l (2 l + 1)\, \frac{2^\frac{3r-2}{2}\, (\pi x)^\frac{3r-5}{2}}{g^{r-2}\, m^{3r-4}}\,,
\end{equation}
where $g$ stands for the internal degrees of freedom of the DM.

Therefore, for $r=2$ and $l=0$ ($s$-wave), the general expression of the thermally-averaged cross section (for $2\rightarrow 2$ process) brings us to the familiar result~\cite{Griest:1989wd}
\begin{equation} \label{MaxCross2t2}
	\langle\sigma_{2\to 2} v\rangle_{\text{max}}^{\text{$s$-wave}} = \frac{4 \sqrt{\pi}}{m^2}\sqrt{x}\,.
\end{equation}
Similarly, the maximum value of the thermally averaged $s$-wave annihilation cross section for $3\to 2$ can be written as
\begin{equation} \label{MaxCross3t2}
    \langle\sigma_{3\to 2} v^2\rangle_{\text{max}}^{\text{$s$-wave}} = \frac{8 \sqrt{2}\, \pi^2}{g}\, \frac{x^2}{m^5}\,.
\end{equation}
Note that the maximum inelastic cross section for identical initial-state particles is twice that of the non-identical scenario. However, a multiplicative factor $1/2$ must be added as a symmetry factor when performing the thermal averaging integral in $\langle\sigma_{2\to 2} v\rangle$ for identical particles in the initial state. As a result, both the expressions in Eqs.~\eqref{MaxCross2tr} and~\eqref{MaxCrossrt2} are valid for identical and non-identical initial-state particles. Now, we pause to mention that the expressions in Eqs.~\eqref{MaxCross2t2} and~\eqref{MaxCross3t2} can be used to put a unitarity bound on the DM mass, as will be shown in Section~\ref{sec5}.

%%%%%%%%%%%%%%%%%%%%%%%%%%%%%%%%%%%%%%%%%%%%%%%%%%%%%%%%%
\section{Low-temperature Reheating}\label{sec4}
%%%%%%%%%%%%%%%%%%%%%%%%%%%%%%%%%%%%%%%%%%%%%%%%%%%%%%%%%
In the standard cosmological paradigm, between the end of inflationary reheating and the beginning of BBN at $T = \Tbbn \simeq 4$~MeV~\cite{Sarkar:1995dd, Kawasaki:2000en, Hannestad:2004px, DeBernardis:2008zz, deSalas:2015glj}, the energy density of the universe is dominated by SM radiation with an energy density $\rR$ given by
\begin{equation}
    \rR(T) = \frac{\pi^2}{30}\, \gs(T)\, T^4,
\end{equation}
where $T$ corresponds to the temperature of the SM bath.
It follows that the Hubble expansion rate $H$ is therefore
\begin{equation}
    H(T) = H_R(T) \equiv \sqrt{\frac{\rR}{3\, M_P^2}} = \frac{\pi}{3}\, \sqrt{\frac{\gs(T)}{10}}\, \frac{T^2}{M_P}\,,
\end{equation}
with $M_P \simeq 2.4 \times 10^{18}$~GeV being the reduced Planck mass.
Additionally, the conservation of the entropy $S \equiv s\, a^3$ of the thermal bath, with $a$ being the cosmic scale factor, and
\begin{equation}
    s(T) = \frac{2\pi^2}{45}\, \gss(T)\, T^3
\end{equation}
the SM entropy density implies that the temperature of the SM bath scales as
\begin{equation} \label{eq:T}
    T(a) \propto \frac{1}{\gss(T)^{1/3}}\, \frac{1}{a}\,.
\end{equation}
Here, $\gs$ and $\gss$ are the numbers of relativistic degrees of freedom that contribute to the energy density of the SM and the entropy of the SM, respectively~\cite{Drees:2015exa}.

However, it is interesting to emphasize that the standard cosmological scenario is not guaranteed and that alternative cosmologies could also have occurred~\cite{Allahverdi:2020bys}. In the following, we focus on cases characterized by low-temperature reheating.\footnote{Note that during our study, we assumed instantaneous thermalization, for details of the effects of non-instantaneous thermalization, see Refs.~\cite{Mukaida:2015ria, Garcia:2018wtq, Chowdhury:2023jft}.} This reheating could correspond to the period just after the end of inflation, or to a secondary period in which an extra component beyond SM radiation dominated the energy density of the universe. In particular, two scenarios will be reviewed: one where the extra component $\phi$ that dominated the expansion of the universe has an energy density that gets diluted faster than radiation and does not decay (that is, a kination-like scenario), and the other where $\phi$ scales as non-relativistic matter and decays into SM particles (that is, an early matter-dominated scenario). These two scenarios will be described below.

%%%%%%%%%%%%%%%%%%%%%%%%%%%%%%%%%%%%%%%%%%%%%%%%%%%%%%%%%
\subsection{Kination-like}
%%%%%%%%%%%%%%%%%%%%%%%%%%%%%%%%%%%%%%%%%%%%%%%%%%%%%%%%%
In this scenario, the universe was dominated by a component $\phi$ whose energy density redshifts faster than free radiation~\cite{DEramo:2017gpl}, as
\begin{equation}
    \rp(a) \propto a^{-(4+n)},
\end{equation}
with $n > 0$.
A typical example of this scenario corresponds to kination~\cite{Spokoiny:1993kt, Ferreira:1997hj}, where $n = 2$.
However, larger values for $n$ are also possible, appearing, for example, in the context of ekpyrotic~\cite{Khoury:2001wf, Khoury:2003rt} or cyclic scenarios~\cite{Gasperini:2002bn, Erickson:2003zm, Barrow:2010rx, Ijjas:2019pyf}; see also Ref.~\cite{Scherrer:2022nnz}.

Interestingly, as this component naturally tends to become subdominant, it is not mandatory to enforce its decay. Let us call $\Trh$ the SM bath temperature at which equality $\rR(\Trh) = \rp(\Trh)$ occurs and from which the universe is dominated by SM radiation (that is, the standard cosmological scenario is recovered).
The Hubble expansion rate is therefore 
\begin{equation} \label{eq:HFast}
    H(T) \simeq H_R(T) \times
    \begin{dcases}
        \left(\frac{T}{\Trh}\right)^{n/2} & \text{ for } T \geq \Trh,\\
        1 & \text{ for } T \leq \Trh,
    \end{dcases}
\end{equation}
where we have taken into account that, as $\phi$ is not decaying, the SM entropy is conserved, and therefore the SM temperature follows the standard scaling shown in Eq.~\eqref{eq:T}.

%%%%%%%%%%%%%%%%%%%%%%%%%%%%%%%%%%%%%%%%%%%%%%%%%%%%%%%%%
\subsection{Early matter domination}
%%%%%%%%%%%%%%%%%%%%%%%%%%%%%%%%%%%%%%%%%%%%%%%%%%%%%%%%%
Alternatively, the universe could have being dominated instead by a component $\phi$ with an energy density that scales like non-relativistic matter: $\rho_\phi(a) \propto a^{-3}$.
As this component tends to dominate the total energy density of the universe, it has to decay.
Again, $\Trh$ is the bath temperature when the equality $\rR(\Trh) = \rp(\Trh)$ occurs, defining the beginning of the SM radiation dominance era.%
\footnote{It is worth mentioning that, in general, the non-adiabatic period could have been preceded by an adiabatic period with $\phi$-domination, and by another era dominated by SM radiation. Here, however, we assume that the DM freeze-out happens in the non-adiabatic period so that previous stages of the universe play no role. This is true if $\phi$ is identified with the inflaton or simply if the nonadiabatic era is long enough.}
Taking into account that the decay of $\phi$ gives rise to a nonadiabatic epoch, the SM temperature scales as~\cite{Giudice:2000ex, Arias:2019uol, Arias:2021rer}
\begin{equation}
    T(a) \simeq \Trh \times
    \begin{dcases}
        \left(\frac{\arh}{a}\right)^{3/8} & \text{ for } T \geq \Trh,\\
        \left(\frac{\gss(\Trh)}{\gss(T)}\right)^{1/3} \frac{\arh}{a} & \text{ for } T \leq \Trh.
    \end{dcases}
\end{equation}
It follows that the Hubble expansion rate is
\begin{equation} \label{eq:HSlow}
    H(T) \simeq
    \begin{dcases}
        H_R(\Trh) \left(\frac{T}{\Trh}\right)^4 & \text{ for } T \geq \Trh,\\
        H_R(T) & \text{ for } T \leq \Trh.
    \end{dcases}
\end{equation}
Interesting, the SM thermal bath can reach temperatures higher than $\Trh$, up to $T = T_\text{max}$~\cite{Giudice:2000ex}.

Having settled the evolution of the background, in the next section the dynamics of the thermal DM in such alternative cosmological scenarios, and in particular the impact on the unitarity limit, will be carefully studied.

%%%%%%%%%%%%%%%%%%%%%%%%%%%%%%%%%%%%%%%%%%%%%%%%%%%%%%%%%
\section{Freeze-out with a Low-temperature Reheating} \label{sec5}
%%%%%%%%%%%%%%%%%%%%%%%%%%%%%%%%%%%%%%%%%%%%%%%%%%%%%%%%%
In this section, two cases for the DM freeze-out are considered. The first corresponds to the visible freeze-out, where a couple of DM particles annihilate into a couple of SM states, with a total thermally-averaged annihilation cross section $\sv$.
The evolution of the DM number density $n$ can be described with the Boltzmann equation~\cite{Lee:1977ua}
\begin{equation} \label{eq:BE10}
    \frac{dn}{dt} + 3\, H\, n = -\sv \left(n^2 - n_\text{eq}^2\right),
\end{equation}
where the DM number density at equilibrium $n_\text{eq}$ is
\begin{equation}
    n_\text{eq}(T) = g \left(\frac{m\, T}{2 \pi}\right)^{3/2} e^{- \frac{m}{T}}
\end{equation}
for non-relativistic DM particles, with $g$ and $m$ being the number of internal degrees of freedom and the mass of the DM particle, respectively.
Alternatively, DM could have been generated through a dark freeze-out (that is, a cannibalization process), where $r$ DM particles annihilate into two DM particles with an interaction given by $\svk$.
In that case, the evolution of $n$ is given by~\cite{Hochberg:2014dra, Bernal:2015ova, Bernal:2015xba}
\begin{equation} \label{eq:BE20}
    \frac{dn}{dt} + 3\, H\, n = -(\Delta n)\svk \left(n^r - n^2\, n_\text{eq}^{r-2}\right).
\end{equation}
where, $\Delta n$ represents the net change of the DM number in the reaction. It is important to recall that in the present analysis, $s$-wave annihilations are always considered, which implies that $\sv$ and $\svk$ are temperature-independent quantities. 

The freeze-out temperature $\Tfo$ corresponds to the temperature at which DM exits chemical equilibrium. It is important to note that here we assume that the kinetic equilibrium between the dark and visible sectors is maintained at least until the end of chemical freeze-out, so that at freeze-out the two sectors share the same temperature. The freeze-out temperature can be estimated by the equality between the Hubble and the interaction rates
\begin{equation} \label{eq:fo1}
    H(\Tfo) = n_\text{eq}(\Tfo)\, \sv,
\end{equation}
or
\begin{equation} \label{eq:fo2}
    H(\Tfo) = n_\text{eq}^{r-1}(\Tfo)\, \svk,
\end{equation}
in the case of a visible or dark freeze-out, respectively.

%%%%%%%%%%%%%%%%%%%%%%%%%%%%%%%%%%%%%%%%%%%%%%%%%%%%%%%%%
\subsection{Kination-like}
%%%%%%%%%%%%%%%%%%%%%%%%%%%%%%%%%%%%%%%%%%%%%%%%%%%%%%%%%
In this case, as the SM entropy is conserved, it is convenient to rewrite Eqs.~\eqref{eq:BE10} and~\eqref{eq:BE20} in terms of the comoving yield $Y(T) \equiv n(T)/s(T)$ and $x \equiv m/T$ as
\begin{equation} \label{eq:BE11}
    \frac{dY}{dx} = - \frac{\sv\, s}{x\, H} \left[Y^2 - Y_\text{eq}^2\right]
\end{equation}
and
\begin{equation} \label{eq:BE21}
    \frac{dY}{dx} = - \frac{ (\Delta n) \svk\, s^{r-1}}{x\, H} \left[Y^r - Y_\text{eq}^{r-2} Y^2\right],
\end{equation}
where $Y_\text{eq} \equiv n_\text{eq}/s$.

The freeze-out temperature $\Tfo = m/\xfo$ can be estimated by comparing Eqs.~\eqref{eq:HFast} with~\eqref{eq:fo1} or~\eqref{eq:fo2}, and can be expressed in the convenient way
\begin{equation} \label{eq:xfoFast}
    \xfo =
    \begin{dcases}
        \frac12\, \frac{3r-n-7}{r-1}\, \mathcal{W}_{-1}\left[2\, \frac{r-1}{3r-n-7} \left(\frac{2^{3r-4}\, g^{2-2r}\, \gs\, \pi^{3r-1}}{45} \frac{\xrh^n\, m^{10-6r}}{M_P^2\, \svk^2}\right)^\frac{1}{7-3r+n}\right]\\
        \hspace{11.5cm} \text{for } 3\,r-n\ne7,\\
        \frac12\, \frac{1}{r-1} \log\left[\frac{45}{2^{3r-4}\, \pi^{3r-1}}\, \frac{g^{2r-2}}{\gs}\, \frac{m^{6r-10}\, \svk^2\, M_P^2}{\xrh^{3r-7}}\right] \hspace{2.0cm} \text{for } 3\,r-n=7,
    \end{dcases}
\end{equation}
with $\xrh \equiv m/\Trh$, and where $\mathcal{W}_{-1}$ correspond to the branch $-1$ of the Lambert $\mathcal{W}$ function. In the case of a visible freeze-out $r = 2$, one has to use that $\svk = \sv$, while in the radiation-dominated era $n = 0$.

In the following, Eqs.~\eqref{eq:BE11} and~\eqref{eq:BE21} will be analytically solved in the context of a kination-like cosmology. For convenience, we start with the case corresponding to the dark freeze-out.

%%%%%%%%%%%%%%%%%%%%%%%%%%%%%%%%%%%%%%%%%%%%%%%%%%%%%%%%
\subsubsection{Dark freeze-out}
%%%%%%%%%%%%%%%%%%%%%%%%%%%%%%%%%%%%%%%%%%%%%%%%%%%%%%%%
If the freeze-out occurs during the radiation-dominated era, Eq.~\eqref{eq:BE21} can be analytically solved, from the DM freeze-out until today (i.e., small temperature and therefore large $x$)
\begin{equation}
    \int_{Y_\text{fo}}^{Y_0} \frac{dY}{Y^r} \simeq \frac{Y_0^{1-r}}{1-r} \simeq - \svk \int_{\xfo}^\infty \frac{s^{r-1}}{x\, H}\, dx\,,
\end{equation}
where $Y_0$ is the DM yield at late times, and we have used the fact that $Y_\text{fo} \gg Y_0$. It follows that the cross section $\svk$ is
\begin{equation} \label{eq:svkSC}
    \svk \simeq 2^{\frac12 - r} \left(\frac{3}{\pi}\right)^{2r-3} 5^{r-\frac32}\, \frac{3r-5}{r-1}\, \frac{\sqrt{\gs}}{\gss^{r-1}}\, \frac{\xfo^{3r-5}}{M_P\, m^{2r-4}\, (m\, Y_0)^{r-1}}\,.
\end{equation}
To match the whole observed DM relic density, it is required that
\begin{equation}
    m\, Y_0\, = \Omega h^2 \, \frac{1}{s_0}\,\frac{\rho_c}{h^2} \simeq 4.3 \times 10^{-10}~\text{GeV},
\end{equation}
with $\rho_c \simeq 1.05 \times 10^{-5}\, h^2$~GeV/cm$^3$ being the critical energy density, $s_0\simeq 2.69 \times 10^3$~cm$^{-3}$ the present entropy density~\cite{ParticleDataGroup:2020ssz}, and $\Omega h^2 \simeq 0.12$ the observed DM relic abundance~\cite{Planck:2018vyg}.

Alternatively, if the freeze-out happens during reheating
\begin{equation}
    \int_{Y_\text{fo}}^{Y_0} \frac{dY}{Y^r} \simeq \frac{Y_0^{1-r}}{1-r} \simeq - \svk \left[\int_{\xfo}^{\xrh} \frac{s^{r-1}}{x\, H}\, dx + \int_{\xrh}^\infty \frac{s^{r-1}}{x\, H}\, dx\right];
\end{equation}
the integral has been split into two pieces, to emphasize the two regimes of $H$ in Eq.~\eqref{eq:HFast}.
Therefore
\begin{align} \label{eq:svkNSC}
    \svk &\simeq 2^{\frac12 - r} \left(\frac{3}{\pi}\right)^{2r-3} 5^{r-\frac32}\, \frac{3r-5}{r-1}\, \frac{\sqrt{\gs}}{\gss^{r-1}}\, \frac{\xfo^{3r-5}}{M_P\, m^{2r-4}\, (m\, Y_0)^{r-1}}\nonumber\\
    &\quad\quad \times \left(\frac{\xrh}{\xfo}\right)^{3r-5} \times
    \begin{dcases}
         \frac{1}{1 + 2\, \frac{3r-5}{6r-n-10} \left[\left(\frac{\xrh}{\xfo}\right)^{3r-n/2-5}-1\right]} &\text{ for } 6\,r-n \ne 10,\\
         \frac{1}{{1 + (3r-5)}\, \log \frac{\xrh}{\xfo}} &\text{ for } 6\,r-n = 10.
    \end{dcases}
\end{align}
As expected, Eq.~\eqref{eq:svkSC} can be recovered from Eq.~\eqref{eq:svkNSC} by taking $n = 0$.

%%%%%%%%%%%%%%%%%%%%%%%%%%%%%%%%%%%%%%%%%%%%%%%%%%%%%%%%
\subsubsection{Visible freeze-out}
%%%%%%%%%%%%%%%%%%%%%%%%%%%%%%%%%%%%%%%%%%%%%%%%%%%%%%%%
The case of the visible freeze-out in Eq.~\eqref{eq:BE11} can be computed following the same procedure presented in the previous subsection. However, one could also derive it by fixing $r = 2$ in Eqs.~\eqref{eq:svkSC} and~\eqref{eq:svkNSC}, which gives
\begin{equation} \label{eq:svrd}
    \sv \simeq \frac{3}{2 \pi} \sqrt{\frac52}\, \frac{\sqrt{\gs}}{\gss}\, \frac{\xfo}{M_P\, m\, Y_0}\,,
\end{equation}
for the freeze-out in the radiation-dominated era, or
\begin{equation} \label{eq:svnsc}
    \sv \simeq \frac{3}{2 \pi} \sqrt{\frac52}\, \frac{\sqrt{\gs}}{\gss}\, \frac{\xfo}{M_P\, m\, Y_0}\, \frac{\xrh}{\xfo} \times
    \begin{dcases}
        \frac{1}{1 + \frac{2}{2-n} \left[\left(\frac{\xrh}{\xfo}\right)^{1-n/2} - 1\right]} &\text{ for } n \ne 2,\\
        \frac{1}{1+ \log \frac{\xrh}{\xfo}} &\text{ for } n = 2,
    \end{dcases}
\end{equation}
during reheating.\\

%%%%%%%%%%%%%%%%%%%%%%%%%%%%%%%%%%%%%%%%%%%%%%%%%%%
\begin{figure}[t!]
    \centering
    \includegraphics[scale=0.5]{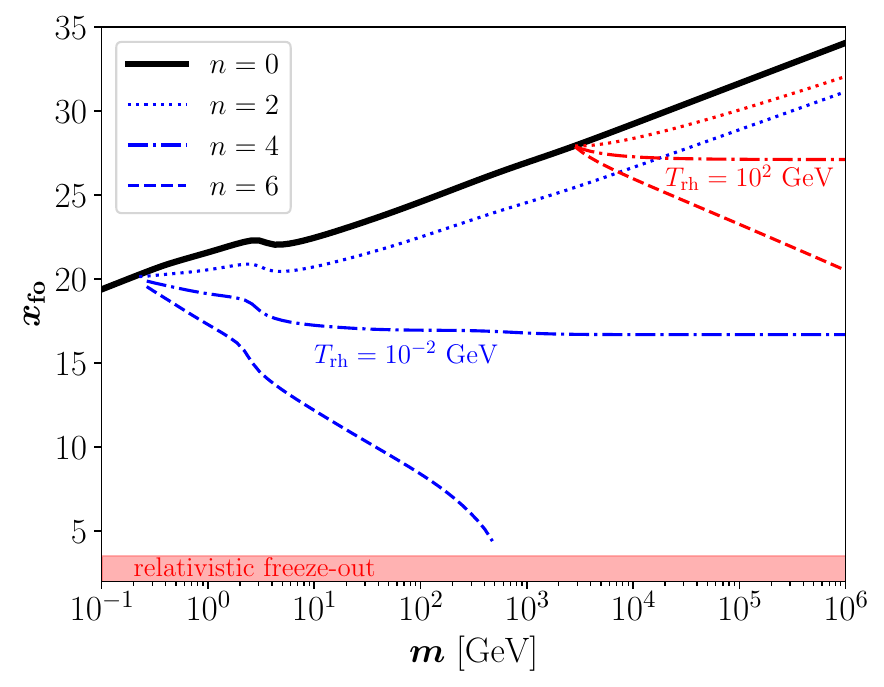}
    \includegraphics[scale=0.5]{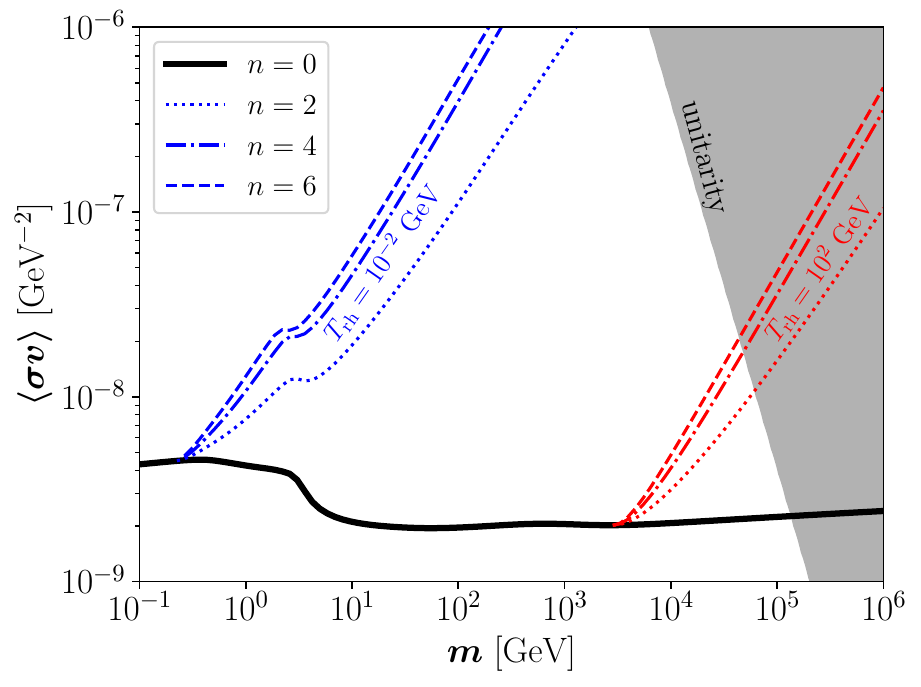}
    \caption{Kination-like. Required freeze-out temperature $\Tfo = m/\xfo$ (left) and cross section $\sv$ (right) for $2 \to 2$ annihilations to fit the observed abundance of DM, in the case of radiation domination ($n=0$) and low-temperature reheating with $n=2$, 4 and 6.
    The blue curves correspond to $\Trh = 10^{-2}$~GeV, while the red curves correspond to $\Trh = 10^2$~GeV. In the red band the freeze-out is relativistic, whereas in the gray band unitarity is violated.}
    \label{fig:Faster_2-to-2a}
\end{figure} 
%%%%%%%%%%%%%%%%%%%%%%%%%%%%%%%%%%%%%%%%%%%%%%%%%%%
The required freeze-out temperature and thermally-averaged annihilation cross section to fit the observed DM abundance, in the case of standard cosmology ($n=0$) and a kination-like scenario with $n=2$, 4 and 6, are shown in Fig.~\ref{fig:Faster_2-to-2a}. The blue curves correspond to $\Trh = 10^{-2}$~GeV, while the red curves correspond to $\Trh = 10^2$~GeV.
In the case of radiation domination, the usual results are recovered: $\xfo \sim \mathcal{O}(25)$ and $\sv \sim \mathcal{O}(10^{-9})$~GeV$^{-2}$, which correspond to a few picobarns, for WIMPs in the GeV ballpark~\cite{Steigman:2012nb}.
Alternatively, in kination-like scenarios, the universe expands faster, and therefore an early freeze-out is required.
High values of $n$ induce small values of $\xfo$, and a potential relativistic freeze-out (i.e. $\xfo \lesssim 3$), depicted as a red band in the left panel.
Additionally, larger cross sections $\sv$ are also required, potentially in tension with the unitarity constraint; cf. Eq.~\eqref{MaxCross2t2}, shown with a gray band in the right panel. It is worth mentioning that, instead of a single constraint, there is a constraint per cosmological scenario, in this case for each choice of $n$ and $\Trh$. However, in Fig.~\ref{fig:Faster_2-to-2a} all the different constraints basically overlap.

%%%%%%%%%%%%%%%%%%%%%%%%%%%%%%%%%%%%%%%%%%%%%%%%%%%
\begin{figure}[t!]
    \centering
    \includegraphics[scale=0.5]{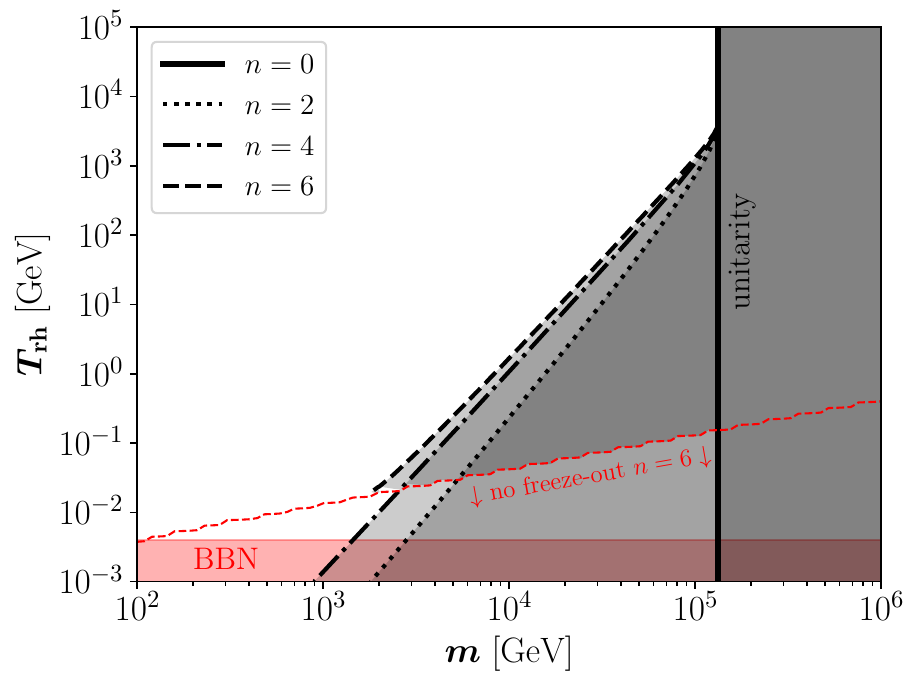}
    \caption{Kination-like. Unitarity bound (gray shaded regions) in the $[m,\, \Trh]$ plane for $2 \to 2$ annihilations. The solid vertical black line corresponds to a freeze-out during radiation-domination ($n=0$), while the black dotted, dot-dashed, and dashed lines correspond to freeze-out during reheating with $n= 2$, 4 and 6, respectively. In the region below the red dashed line, there is no freeze-out for $n=6$.
    In the red band, reheating occurs after BBN.}
    \label{fig:Faster_2-to-2b}
\end{figure} 
%%%%%%%%%%%%%%%%%%%%%%%%%%%%%%%%%%%%%%%%%%%%%%%%%%%
The impact of the unitarity constraint in the case of a $2 \to 2$ scattering becomes more clear in Fig.~\ref{fig:Faster_2-to-2b}, where the gray bands show the excluded regions by unitarity following Eq.~\eqref{MaxCross2t2}, for $n=0$, 2, 4 and 6, in the plane $[m,\, \Trh]$.
The standard upper bound of $\sim 1.3 \times 10^5$~GeV for the DM mass becomes tighter and depends on $\Trh$, if the freeze-out happens in a kination-like era, as expected from Fig.~\ref{fig:Faster_2-to-2a}.
For example, if $\Trh = \Tbbn$, the upper bound on the DM mass can be as strong as $\sim 3 \times 10^3$~GeV or $\sim 1.2 \times 10^3$~GeV for $n=2$ and $n=4$, respectively.
Interestingly, for the case $n=6$, DM may not reach chemical equilibrium in very low-temperature reheating scenarios and therefore freeze-out cannot occur, corresponding to the red dashed line.
Finally, the red band corresponding to $\Trh < \Tbbn$ is in tension with BBN.

%%%%%%%%%%%%%%%%%%%%%%%%%%%%%%%%%%%%%%%%%%%%%%%%%%%
\begin{figure}[t!]
    \centering
    \includegraphics[scale=0.5]{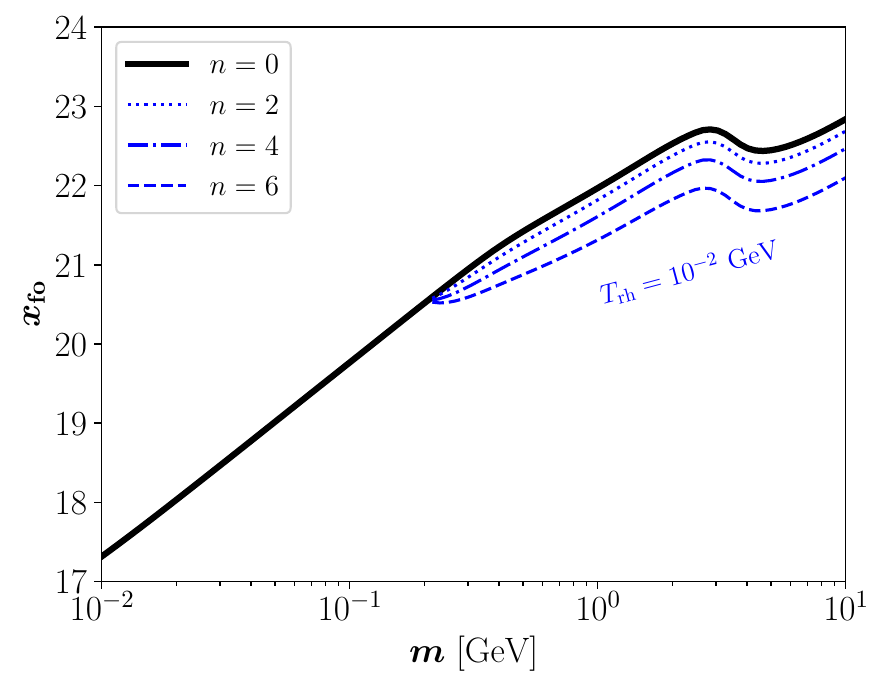}
    \includegraphics[scale=0.5]{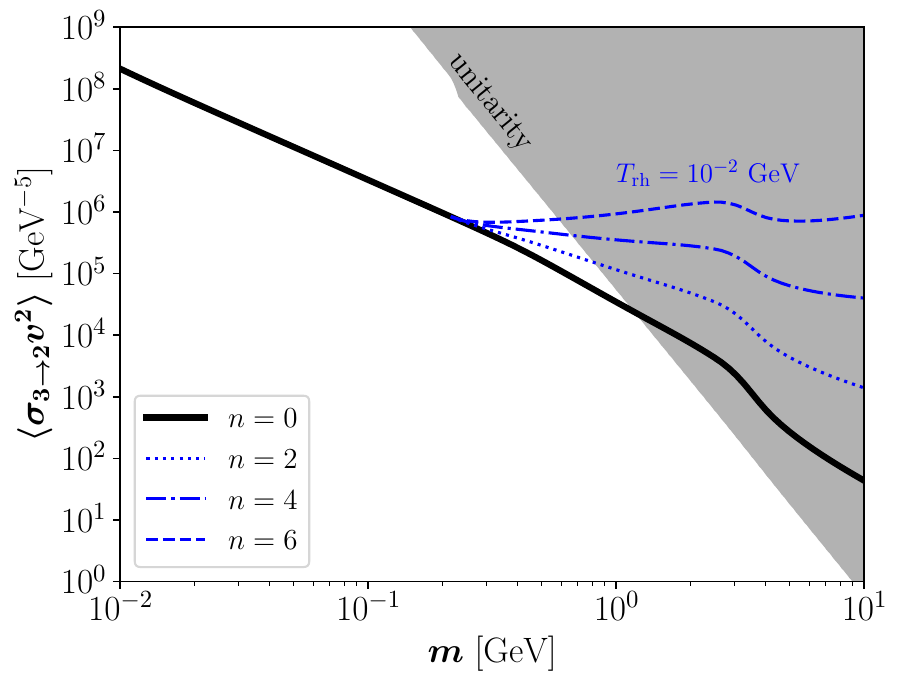}
    \includegraphics[scale=0.5]{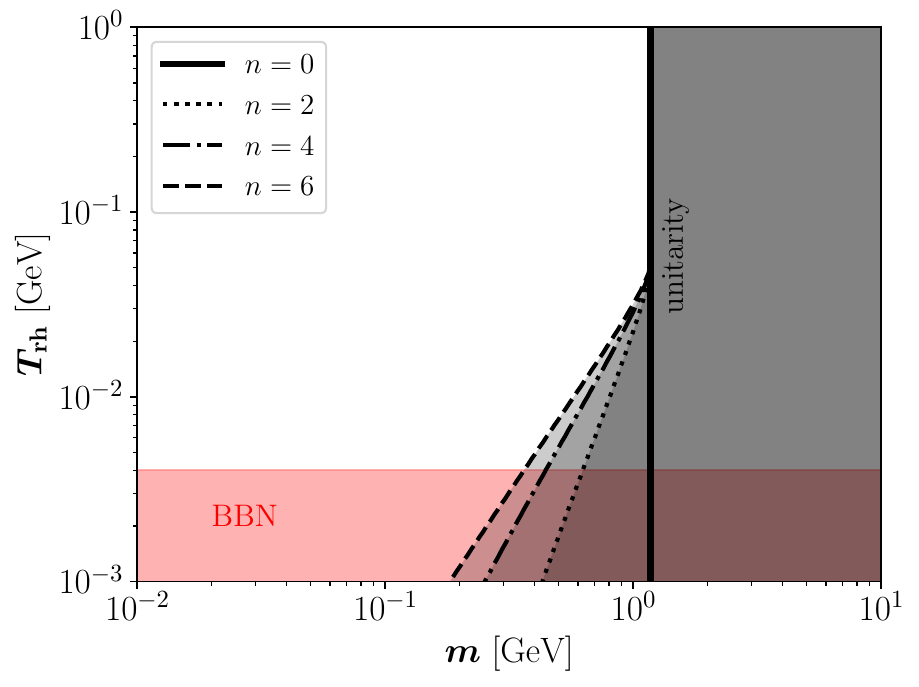}
    \caption{Kination-like. The same as Figs.~\ref{fig:Faster_2-to-2a} and~\ref{fig:Faster_2-to-2b}, but for dark freeze-out through $3 \to 2$ annihilations.}
    \label{fig:Faster_3-to-2}
\end{figure}
%%%%%%%%%%%%%%%%%%%%%%%%%%%%%%%%%%%%%%%%%%%%%%%%%%%
Additionally, the details for the dark freeze-out through $3 \to 2$ scatterings and the unitarity bound following Eq.~\eqref{MaxCross3t2} are shown in Fig.~\ref{fig:Faster_3-to-2}.
In this case, smaller DM masses, in the MeV ballpark, are required. Therefore, the impact of low-temperature reheating is much milder, once the BBN bound on $\Trh$ is imposed.
In a radiation-dominated scenario, the unitarity bound implies $m \lesssim 1$~GeV; however, if freeze out occurs in a kination-like epoch, the bound becomes more stringent, implying that if $\Trh = \Tbbn$, $m \lesssim 7 \times 10^{-1}$~GeV, $m \lesssim 5 \times 10^{-1}$~GeV, or even $m \lesssim 4 \times 10^{-1}$~GeV, for the cases with $n=2$, $n=4$ and $n=6$, respectively.
We note that the most stringent limits on DM mass occur for the kination-like scenario for $\Trh = \Tbbn$ and large values of $n$.

%%%%%%%%%%%%%%%%%%%%%%%%%%%%%%%%%%%%%%%%%%%%%%%%%%%%%%%%%
\subsection{Early matter domination}
%%%%%%%%%%%%%%%%%%%%%%%%%%%%%%%%%%%%%%%%%%%%%%%%%%%%%%%%%
In this case, in order to have a successful reheating, $\phi$ has to decay into SM particles and hence the SM entropy is not conserved.\footnote{It is important to emphasize that we are assuming that DM is only produced from the scattering of SM particles. In particular, we disregard the possible direct production from decays of $\phi$. This is typically a good assumption, as long as its branching fraction into DM particles is smaller than $\sim 10^{-4} \times m/(100~\text{GeV})$~\cite{Drees:2017iod, Arias:2019uol}.}
Therefore, instead of the yield $Y$, it is convenient to use the comoving DM number density $N(a) \equiv n(a) \times a^3$.
Equations~\eqref{eq:BE10} and~\eqref{eq:BE20} can be rewritten, respectively, as
\begin{equation} \label{eq:BE12}
    \frac{dN}{da} = - \frac{\sv}{a^4\, H} \left[N^2 - N_\text{eq}^2\right]
\end{equation}
and
\begin{equation} \label{eq:BE22}
    \frac{dN}{da} = - \frac{\svk}{a^{3r-2}\, H} \left[N^r - N_\text{eq}^{r-2} N^2\right]
\end{equation}
with $N_\text{eq} \equiv n_\text{eq} \times a^3$.

The freeze-out temperature can be estimated by comparing Eqs.~\eqref{eq:HSlow} and~\eqref{eq:fo1} or~\eqref{eq:fo2}, and is given by
\begin{equation} \label{eq:xfoSlow}
    \xfo =
    \begin{dcases}
        \frac12\, \frac{3r-7}{r-1}\, \mathcal{W}_{-1}\left[2\, \frac{r-1}{3r-7} \left(\frac{2^{3r-4}\, g^{2-2r}\, \gs\, \pi^{3r-1}}{45} \frac{m^{10-6r}}{M_P^2\, \svk^2}\right)^\frac{1}{7-3r}\right]\\
        \hspace{10.4cm} \text{for } \xfo \gg \xrh,\\
        \frac12\, \frac{3r-11}{r-1}\, \mathcal{W}_{-1}\left[2\, \frac{r-1}{3r-11} \left(\frac{2^{3r-4}\, g^{2-2r}\, \gs\, \pi^{3r-1}}{45} \frac{\xrh^4\, m^{10-6r}}{M_P^2\, \svk^2}\right)^\frac{1}{11-3r}\right]\\
        \hspace{10.4cm} \text{for } \xfo \ll \xrh,
    \end{dcases}
\end{equation}
where it is always assumed that $T_\text{max} \gg \Tfo$.
Interestingly, Eq.~\eqref{eq:xfoSlow} is also valid for the case of a visible freeze-out, taking $r = 2$ and replacing $\svk$ by $\sv$.
Furthermore and as expected, if the freeze-out occurs in the standard radiation-dominated era, Eq.~\eqref{eq:xfoFast} ($n = 0$) and Eq.~\eqref{eq:xfoSlow} ($\xfo \gg \xrh$), are equivalent.
 
Next, analytical solutions are presented for Eqs.~\eqref{eq:BE12} and~\eqref{eq:BE22} in the context of an early matter-dominated scenario. We will begin with the case corresponding to the dark freeze-out for convenience.

%%%%%%%%%%%%%%%%%%%%%%%%%%%%%%%%%%%%%%%%%%%%%%%%%%%%%%%%
\subsubsection{Dark freeze-out}
%%%%%%%%%%%%%%%%%%%%%%%%%%%%%%%%%%%%%%%%%%%%%%%%%%%%%%%%
If the freeze-out occurs during the radiation era, the solution of Eq.~\eqref{eq:BE22}, or equivalently of Eq.~\eqref{eq:BE21}, is the one presented in Eq.~\eqref{eq:svkSC}.
Instead, if it occurs during the reheating period, one has that
\begin{align}
    \int_{N_\text{fo}}^{N_0}\frac{dN}{N^r} &\simeq \frac{N_0^{1-r}}{1-r} \simeq - \svk \left[\int_{\afo}^{\arh} \frac{da}{a^{3r-2}\, H} + \int_{\arh}^\infty \frac{da}{a^{3r-2}\, H}\right] \nonumber\\
    &\simeq - \svk \left[\frac{2}{9-6r}\, \frac{\arh^{3-3r}}{H(\Trh)} \left(1 - \left(\frac{\afo}{\arh}\right)^{\frac92-3r}\right) + \frac{\arh^{3-3r}}{(3r-5)\, H(\Trh)}\right],
\end{align}
where $N_\text{fo} \equiv N(\afo)$, $N_0$ is the asymptotic value of $N(a)$ at large values of $a$, much after the freeze-out, and we have used the fact that $N_\text{fo} \gg N_0$.
Here again, the integral has been spit into the two regimes of $H$ in Eq.~\eqref{eq:HSlow}.
The thermally-averaged cross-section is, therefore,
\begin{equation} \label{eq:svkSlow}
    \svk \simeq \frac{5^{r-\frac32} 9^{r-1}}{2^{r-\frac12} \pi^{2r-3}} \frac{(2r-3) (3r-5)}{r-1} \frac{\sqrt{\gs}}{\gss^{r-1}} \frac{\xrh^{3r-5}}{1 + 2(3r-5) \left(\frac{\xrh}{\xfo}\right)^{8r-12}} \frac{m^{4-2r}}{M_P\, (m\, Y_0)^{r-1}}\,.
\end{equation}

%%%%%%%%%%%%%%%%%%%%%%%%%%%%%%%%%%%%%%%%%%%%%%%%%%%%%%%%
\subsubsection{Visible freeze-out}
%%%%%%%%%%%%%%%%%%%%%%%%%%%%%%%%%%%%%%%%%%%%%%%%%%%%%%%%
If the freeze-out occurs during radiation domination, the solution of Eq.~\eqref{eq:BE12} is the same as the one of Eq.~\eqref{eq:svrd}.
Alternatively, if it occurs during reheating, one has instead
\begin{equation}
    \sv \simeq \frac{9}{2 \pi} \sqrt{\frac52} \frac{\sqrt{\gs}}{\gss}\, \frac{\xrh}{1 + 2 \left(\frac{\xrh}{\xfo}\right)^4}\, \frac{1}{M_P\, m\, Y_0}\,,
\end{equation}
simply corresponding to the limit $r = 2$ of Eq.~\eqref{eq:svkSlow}.\\

%%%%%%%%%%%%%%%%%%%%%%%%%%%%%%%%%%%%%%%%%%%%%%%%%%%
\begin{figure}[t!]
    \centering
    \includegraphics[height=6.3cm,width=7.9cm]{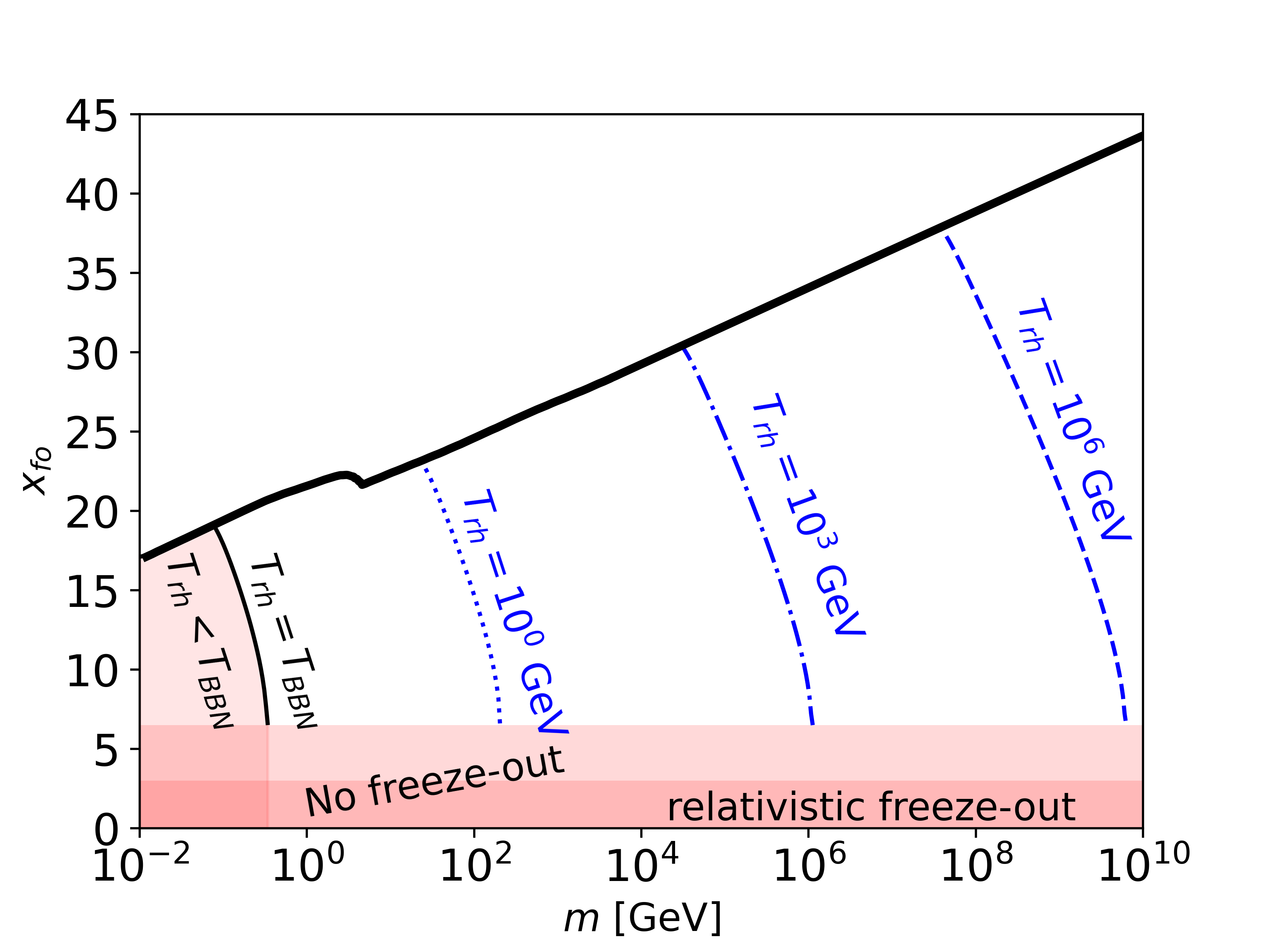}~
    \includegraphics[height=6.3cm,width=7.9cm]{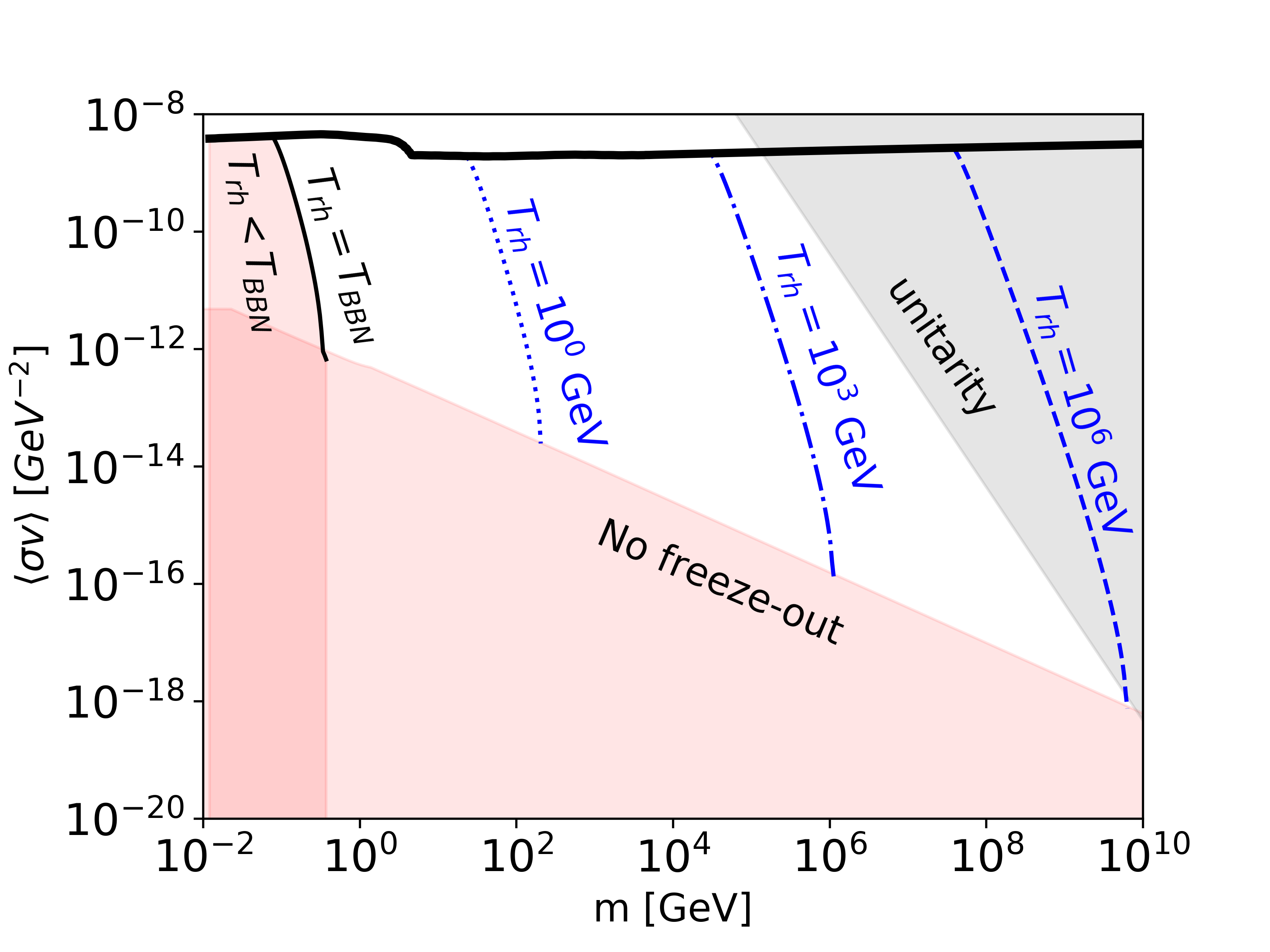}
    \includegraphics[height=6.3cm,width=7.9cm]{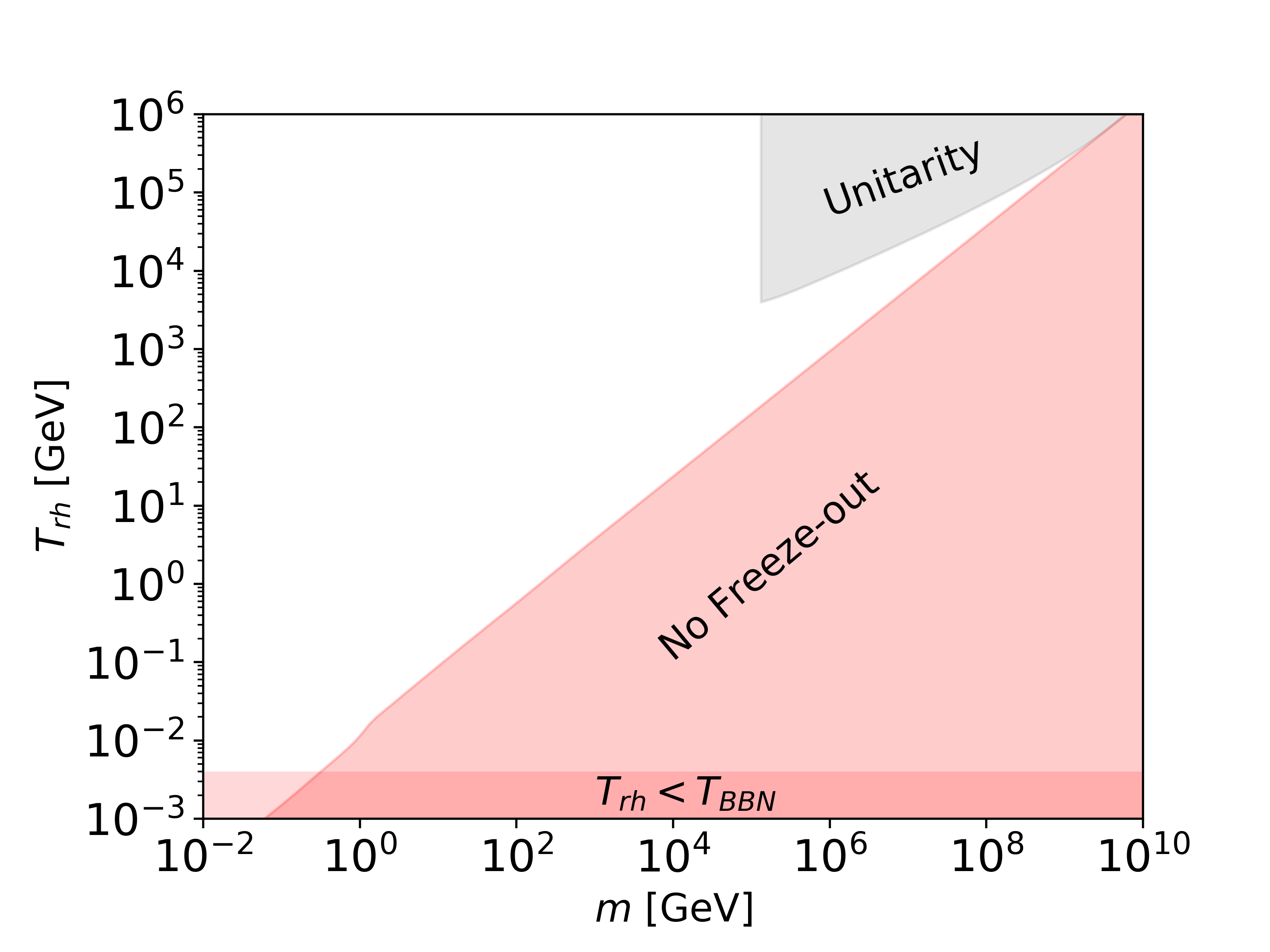}
    \caption{Early matter domination.
    The same as in Figs.~\ref{fig:Faster_2-to-2a} and~\ref{fig:Faster_2-to-2b}, for $2 \to 2$ annihilations, but for early matter domination. Additionally, the ``No freeze-out'' constraint is shown by the red-shaded region.}
    \label{fig:Slower_2-to-2a}
\end{figure} 
%%%%%%%%%%%%%%%%%%%%%%%%%%%%%%%%%%%%%%%%%%%%%%%%%%%
The left panel of Fig.~\ref{fig:Slower_2-to-2a} displays the freeze-out temperature needed to fit the observed relic of DM as a function of its mass for radiation-dominated universe (thick black solid line) and reheating temperatures, $\Trh=\Tbbn$ (black solid lines), $10^0$~GeV (blue dotted line), $10^3$~GeV (blue dot-dashed line), $10^6$~GeV (blue dashed line) where the DM number changing process is $2 \to 2$, that is the visible WIMP mechanism. The dilution by large entropy injection has to be overcome by the overproduction of DM at freeze-out, which results in an earlier freeze-out and therefore a small $\xfo$. Interestingly, chemical equilibration of DM requires $\xfo \gtrsim 6.5$, which corresponds to the red band. 
Smaller values for $\xfo$ could lead to the observed abundance of DM, but not through the WIMP mechanism, but instead through the FIMP paradigm~\cite{McDonald:2001vt, Choi:2005vq, Kusenko:2006rh, Petraki:2007gq, Hall:2009bx, Elahi:2014fsa, Bernal:2017kxu}.
An additional red band corresponds to $\xfo \le 3$ and represents the relativistic freeze-out.
The right panel of Fig.~\ref{fig:Slower_2-to-2a} contains equivalent information, but is now presented in the plane $[m, \sv]$.
High freeze-out temperatures correspond to earlier decouplings and, therefore, to lower cross sections. Again, $\sv$ is bounded from below by the requirement of reaching chemical equilibrium and corresponds to the ``No freeze-out'' region.
It is important to note that the cross section decreases with increasing DM mass for a fixed $\Trh$. The reason is that the freeze-out of heavier DM occurs earlier (that is, at a larger $\Tfo$), demanding a smaller $\sv$.
Additionally, the gray-shaded region corresponding to high values of $\sv$ is disallowed by unitarity following Eq.~\eqref{MaxCross2t2}.\footnote{We note that for a given point $[m,\, \sv]$ one can compute required freeze-out and reheating temperatures required to fit the whole observed DM abundance. The corresponding $\xfo$ is used in Eq.~\eqref{MaxCross2t2} to calculate the unitarity bound.}
In low-temperature reheating scenarios with large injection of entropy, thermally averaged cross sections much smaller and DM masses much larger than the canonical values $\sv \sim \mathcal{O}(10^{-9})$~GeV$^{-2}$ and $m \sim 1.3 \times 10^3$~GeV are allowed.
Interestingly, extreme values of $m$ and $\sv$ compatible with unitarity can be reached for $\Trh \simeq 10^6$~GeV corresponding to $m \simeq 10^{10}$~GeV and $\sv \simeq 10^{-18}$~GeV$^{-2}$.

In the bottom panel of Fig.~\ref{fig:Slower_2-to-2a}, we have shown the allowed parameter space in the reheating temperature and DM mass plane where the white space is free of any constraints. The red band dictates that $\Trh$ lower than $\Tbbn$ is in disagreement with the cosmological observation. The red and gray-shaded region displays the ``No freeze-out'' and the unitarity constraints discussed earlier.
Note that the unitarity constraint is independent of the reheating temperature for $m \sim 1.3 \times 10^5$~GeV, reflected by the straight vertical shape in the unitarity bound. This is the DM mass where the cross section needed to fit the observed DM abundance (in a radiation-dominated background) and the maximum allowed cross section by unitarity match exactly.
However, low-temperature reheating opens up the possibility of higher masses.
Interestingly, masses higher than $m \sim 1.3 \times 10^5$~GeV are severely constrained from above due to unitarity and from below due to the requirement of chemical equilibrium, resulting in a narrow allowed region for $\Trh$.

%%%%%%%%%%%%%%%%%%%%%%%%%%%%%%%%%%%%%%%%%%%%%%%%%%%
\begin{figure}[t!]
    \centering
    \includegraphics[height=6.3cm,width=7.9cm]{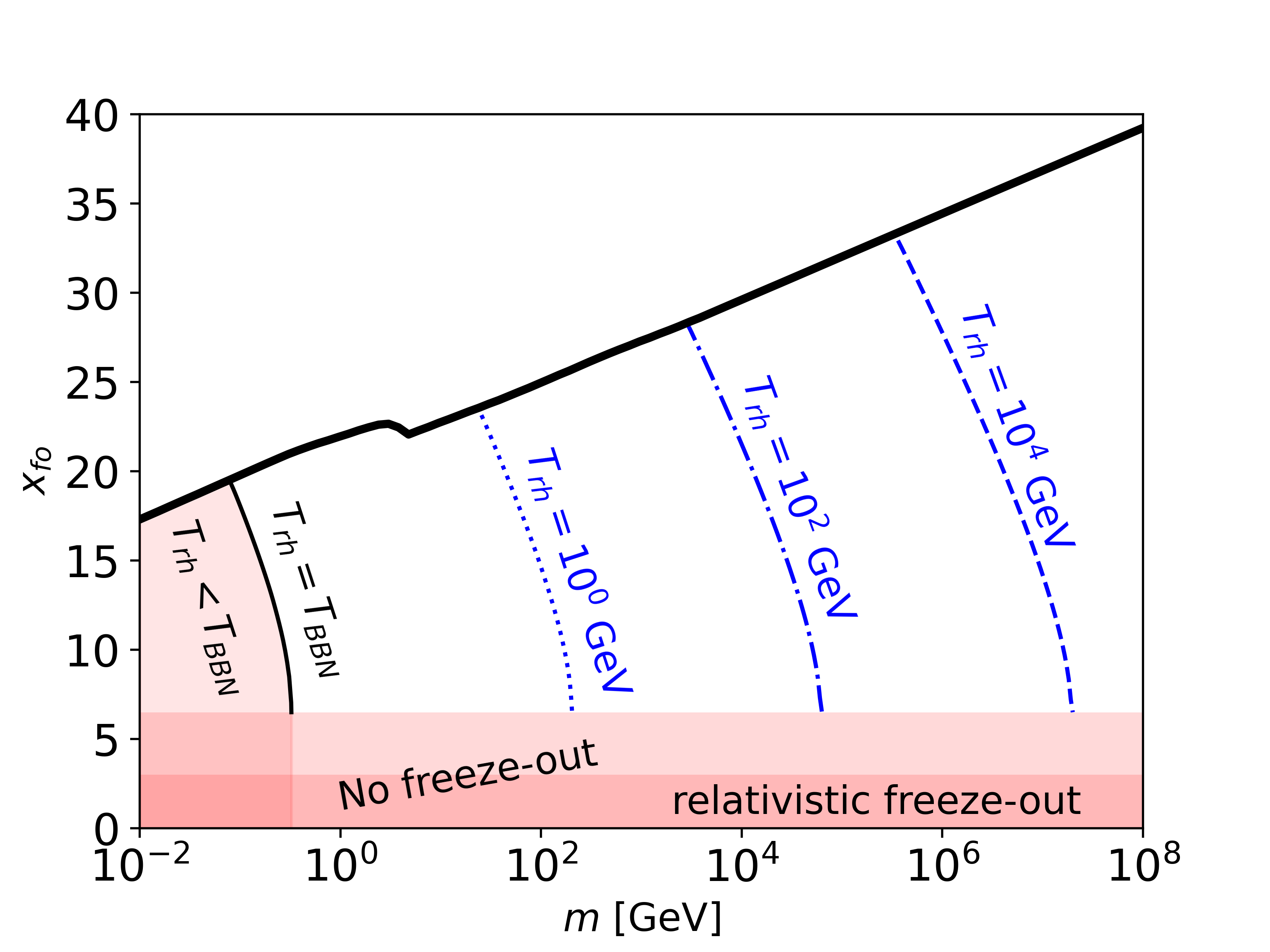}~
    \includegraphics[height=6.3cm,width=7.9cm]{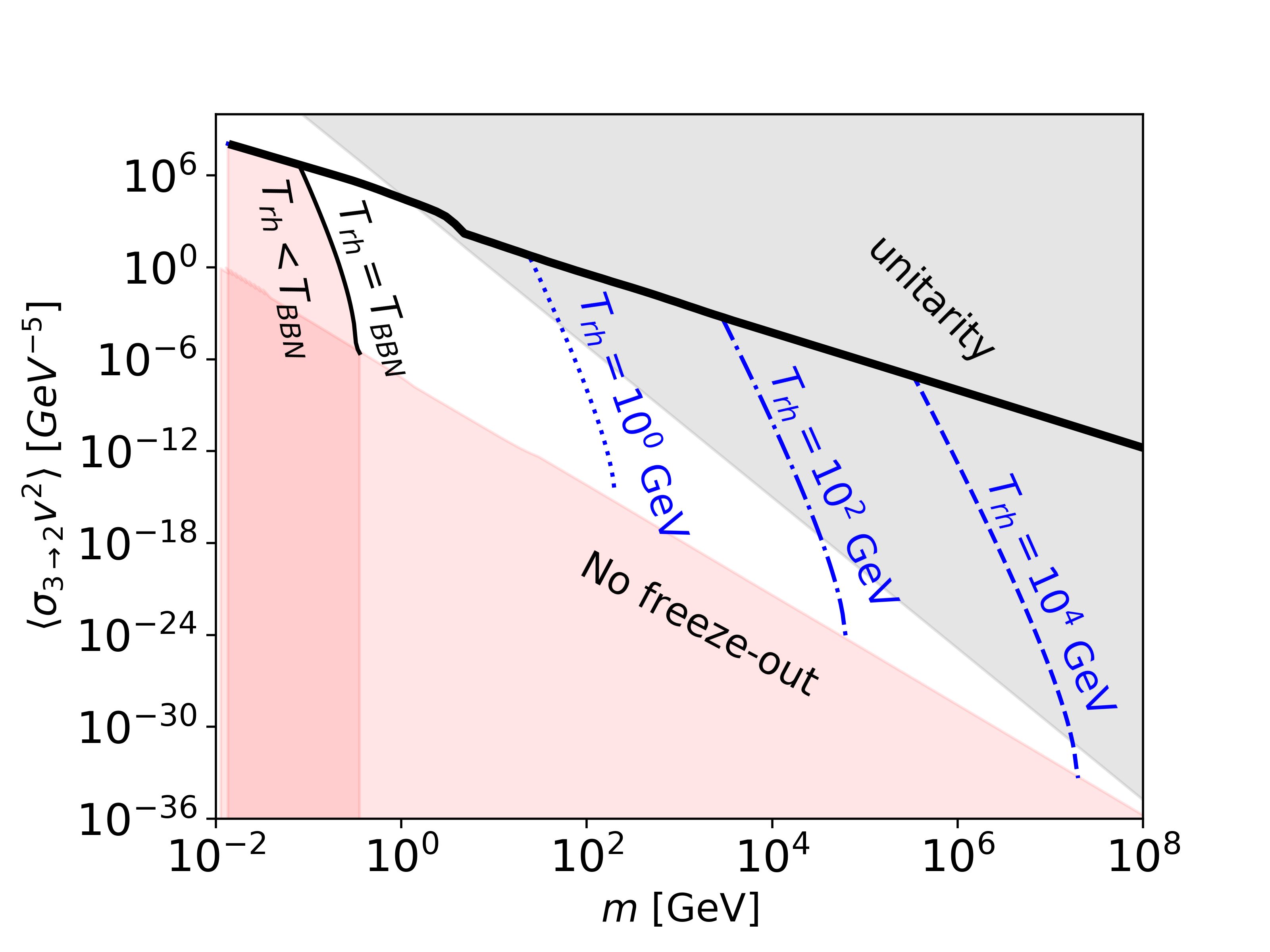}\\
    \includegraphics[height=6.3cm,width=7.9cm]{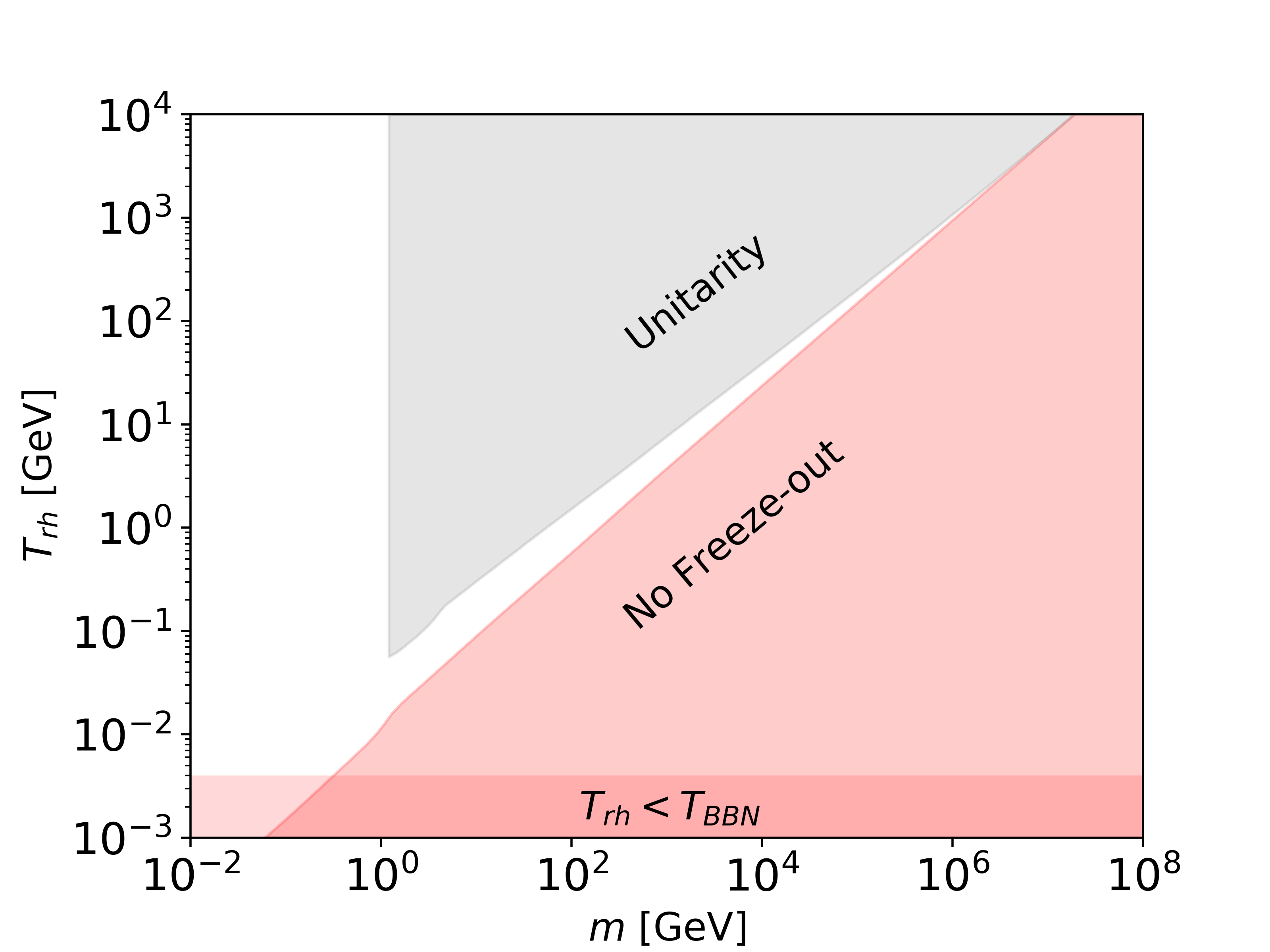}
    \caption{ Early matter domination. The same as in Fig.~\ref{fig:Slower_2-to-2a}, but for dark freeze-out through 3-to-2 annihilations.}
    \label{fig:Slower_3-to-2}
\end{figure} 
%%%%%%%%%%%%%%%%%%%%%%%%%%%%%%%%%%%%%%%%%%%%%%%%%%%
For completeness, Fig.~\ref{fig:Slower_3-to-2} depicts results equivalents to the ones shown in Fig.~\ref{fig:Slower_2-to-2a} but now for $3\to2$ annihilations. In this case, the presence of late-time reheating opens up the DM mass to $\sim 10^8$~GeV with $\langle\sigma_{3\to 2}v^2\rangle \sim 10^{-36}$~GeV$^{-5}$.

%%%%%%%%%%%%%%%%%%%%%%%%%%%%%%%%%%%%%%%%%%%%%%%%%%%%%%%%%
\section{Summary and Conclusion} \label{sec6}
%%%%%%%%%%%%%%%%%%%%%%%%%%%%%%%%%%%%%%%%%%%%%%%%%%%%%%%%%
The requirement of the de Broglie wavelength of dark matter (DM) to hold it inside galaxies and the stability of the stellar cluster in galaxies collectively put a broad allowed mass range for DM by providing the lower and upper bound of DM mass, respectively. Specifying some properties of DM can further tighten the mass range. 
Interestingly, one can put the model-independent upper bound by specifying the thermal production of DM in the early universe. The observed DM abundance and unitarity of partial waves from the scattering matrix jointly place an upper limit on DM mass. Primarily, the upper limits on the inelastic cross section for a general number-changing process $2\to r$ can be derived with the help of the optical theorem, the matrix elements and the elastic scattering cross section for the process. After that, one can obtain the thermally averaged cross section for the $r\to 2$ process by invoking the principle of detailed balance. Finally, the unitarity bounds on the thermally averaged cross section translate to the upper limits on the DM mass satisfying the relic density constraints. It is known that the maximum allowed DM mass for the $2\to 2$ and $3\to 2$ DM annihilation processes is around 130~TeV and 1~GeV, respectively. However, these bounds depend not only on the particle physics model, but also have a strong dependence on the cosmological evolution of the universe, being valid only if the universe followed the so-called ``standard cosmological scenario''.

Instead, this article explores the DM mass bound in non-standard cosmological setups characterized by low-temperature reheating. In particular, we focus on $i)$ kination-like scenarios, where the early universe was dominated by a fluid with an energy density that gets diluted faster than free radiation, and $ii)$ early matter-dominated scenarios, where a component with an energy density that scales as nonrelativistic matter dominates the early universe and eventually decays into SM particles.

First, we study the kination-like universe, which demands a larger thermally-averaged annihilation cross section to saturate the observed abundance of DM compared to the standard radiation-dominated picture since, in this case, freeze-out occurs early. As a result, the upper bound on the DM mass becomes more stringent than in the standard case. For example, if the reheating temperature is as low as a few MeVs (corresponding to the start of the Big Bang nucleosynthesis epoch), the usual bound of the DM mass $m \lesssim 130$~TeV can be reduced to a few TeVs for WIMPs.

Second, we also consider the picture of early matter domination, which dictates a fast expansion with entropy production. Although the DM freezes out early in this scenario, here one needs a smaller thermally averaged cross section to feature an observed relic of DM than a radiation-dominated picture to compensate for the dilution of the relic due to the huge entropy injection. Thus, the presence of an early matter domination relaxes the mass bound, and eventually the allowed mass range of the DM is enhanced. Here, the usual bound of the DM mass $m \lesssim 130$~TeV can be relaxed, making the WIMP DM masses up to $\sim 10^{10}$~GeV viable. 

Before closing, we want to emphasize that the evolution of the early universe is largely unknown. The standard assumption of a universe dominated by standard-model radiation from the end of cosmological inflation until matter-radiation equality, together with a transition from an inflaton-dominated to a radiation-dominated universe occurring at a very early time, cannot be taken for granted. Having that in mind, here we have studied the impact of the unitarity bound on DM in the case of low-temperature reheating scenarios.

%%%%%%%%%%%%%%%%%%%%%%%%%%%%%%%%%%%%%%%%%%%%%%%%%%%%%%%%%
\acknowledgments
%%%%%%%%%%%%%%%%%%%%%%%%%%%%%%%%%%%%%%%%%%%%%%%%%%%%%%%%%
The authors acknowledge the hospitality during the IMHEP 23 at IOP, Bhubaneswar, where this project was initiated.
Computational work was performed on the Param Vikram-1000 High-Performance Computing Cluster and TDP resources at the Physical Research Laboratory (PRL).

%%%%%%%%%%%%%%%%%%%%%%%%%
\bibliographystyle{JHEP}
\bibliography{biblio}
%%%%%%%%%%%%%%%%%%%%%%%%%
\end{document}